\documentclass[apj,twocolumn,appendixfloats,twocolappendix]{openjournal}

\usepackage[T1]{fontenc}
\usepackage[utf8]{inputenc}

\usepackage{amsmath} 
\usepackage{amsfonts} 
\usepackage{amsbsy} 
\usepackage{amssymb} 

\usepackage{graphicx} 

\usepackage{hyperref} 

\usepackage{acronym} 
\usepackage{soul} 
\usepackage{xcolor} 

\usepackage{orcidlink} 

\hypersetup{
    colorlinks=true,
    linkcolor=blue,
    filecolor=blue,
    urlcolor=blue,
    citecolor=blue,
}

\newcommand{\pushright}[1]{\tag*{$#1$}}

\newacro{PPD}{proto-planetary disc}

\newacroplural{PPD}{proto-planetary discs}

\newacro{BC}{boundary condition}

\newacroplural{BC}{boundary conditions}

\newacro{PDE}{partial differential equation}

\newacro{ODE}{ordinary differential equation}

\newcommand{\bb}[1]{\mbox{\boldmath{$#1$}}}

\newcommand{\rd}{\mathrm{d}} 

\newcommand{\bcdot}{\, \mbox{\boldmath{$\cdot$}} \,}

\newcommand{\ex}{\ensuremath{\bb{e}_{x}}} 
\newcommand{\ey}{\ensuremath{\bb{e}_{y}}} 
\newcommand{\er}{\ensuremath{\bb{e}_{r}}} 

\newcommand{\br}{\mathbf{r}} 
\newcommand{\ba}{\mathbf{a}} 

\newcommand{\e}{\mathrm{e}} 

\begin{document}

\title{Reflex instabilities I: Boundaries make grid-based disc simulations frame-dependent}
\shorttitle{Reflex instabilities I}

\author{\vspace{-40pt}Nathan Magnan \orcidlink{0000-0001-8602-9725}$^{1, 2}$}
\author{Michel Tagger$^{3}$}
\author{Clément Baruteau$^{4}$}
\author{Aurélien Crida$^{1}$}
\author{Jean-François Gonzalez \orcidlink{0000-0001-9423-6062}$^{5}$}
\author{Héloïse Méheut$^{1}$}
\affiliation{$^{1}$ Université Côte d'Azur, Observatoire de la Côte d'Azur, CNRS, Laboratoire Lagrange, France}
\affiliation{$^{2}$ Centre National d’\'Etudes Spatiales (CNES), Paris, France}
\affiliation{$^{3}$ LPC2E, OSUC, Observatoire de Paris, Université d'Orléans, Université Paris Sciences et Lettres, CNRS, CNES, Orléans, France}
\affiliation{$^{4}$ IRAP, Université de Toulouse, CNRS, CNES, Toulouse, France}
\affiliation{$^{5}$ Université Lyon 1, ENS de Lyon, CNRS, CRAL, UMR 5574, Saint-Genis-Laval, France
\vspace{+0.5 \baselineskip}
}
\email{Corresponding author: nathan.magnan@oca.eu}

\begin{abstract}
    If the mass distribution within an accretion disc is asymmetric, it will exert a gravitational pull on its central object. The central object must then move away from the origin of the reference frame, resulting in atypical gravitational pulls on the disc. The goal of the present series of papers is to investigate the stability of this feedback loop. In this first paper, we report a surprising finding: the choice of reference frame affects the number and nature of the waves and instabilities contained in the analytical equations. To show this, we use a simple example: the modes discovered by \citet[Appendix~A.3]{Sanchez+25}. We provide the first analytical description of those modes, we explain their physical mechanism, and propose to call them `bracket modes' because they arise from the bracket side of an integration by parts. We show that the same disc can be stable in the stellocentric reference frame but unstable in the barycentric reference frame. This is because the boundaries of the simulation domain are frame-dependent. To avoid this, one must model the entire  disc all the way from the central cavity to the outer edge. But this is prohibitively expensive, and some discs do not have a central cavity anyway. Our results suggest that extreme caution is required when interpreting simulations of asymmetric discs, and that our community may not have the tools to model the reflex motion of the central object yet.
    \vspace{+0.5 \baselineskip}
\end{abstract}

\maketitle

\section{Introduction}
\label{sec:Introduction}

Many accretion discs show asymmetric features. For example, tidal disruption events create ejecta with a shared perigee but different apogees, so they almost certainly create eccentric discs \citep{Rees88}. Circumprimary and circumbinary discs are also often eccentric. The best observational examples of this are the superhumps that plague the light curves of many cataclysmic variables and low-mass X-ray binaries (see \citealt{Fragile+26} and references therein). We also see large-scale spirals and large-scale crescents in protoplanetary discs \citep{Bae+23}. Finally, planets can be seen as extreme asymmetries in the mass distribution.

Whenever one of those asymmetries has a dipolar (\textit{i.e.} ${ m = 1 }$) component, it exerts a gravitational pull on the central object. This displaces the central object and perturbs the gravitational field in the whole system. This perturbation affects the disc's dynamics and mass distribution, creating a closed feedback loop. The present series of papers is concerned with this loop's stability.

In this first paper, we start with a seemingly innocuous aspect of the problem: the choice of reference frame. Specifically, we show that the barycentric and stellocentric reference frames are not equivalent.

To do so, we use a simple example. We study analytically a class of ${m = 1}$ modes discovered recently by \cite{Sanchez+25} and which we call `bracket modes' because they arise from the bracket side of an integration by parts. We will see that their number and stability depend on the choice of reference frame.

Our paper is organised as follows: we introduce the governing equations for both reference frames in \S\ref{sec:Governing_equations}, describe the equilibrium state of the system in \S\ref{sec:Equilibrium}, and derive the linear perturbation equations in \S\ref{sec:Linear_perturbation_equations}. We then build the first analytical description of the bracket modes. We start with the barycentric reference frame in~\S\ref{sec:Bracket_modes_barycentric}, and move on to the stellocentric reference frame in~\S\ref{sec:Bracket_modes_stellocentric}. Finally, we discuss the validity and consequences of our findings in \S\ref{sec:Discussion_conclusion}. 

We invite all readers interested in the physics of the bracket modes to read \S\ref{sec:Barycentric_bracket_modes_mechanism}. There we leverage the analytics of \S\ref{sec:Bracket_modes_barycentric} to uncover the barycentric modes' mechanism.

\section{Governing equations}
\label{sec:Governing_equations}

\newcommand{\Ms}{M_{\star}} 
\newcommand{\Md}{M_{d}} 

We consider a two-dimensional disc of mass $\Md$ around a point-like central object of mass $\Ms$. Bracket modes arise even when there is no pressure, so to keep things simple we assume that the disc is only subject to two forces: the gravity of the central object, and its own gravity. Since there is no pressure, this model is better adapted to debris discs and planetary rings than accretion discs, but we shall see in a following paper that bracket modes can accommodate pressure, so they also affect accretion discs.

We introduce the equations governing the disc's evolution in \S\ref{sub:Governing_equations_disc}, those governing the central object's motion in \S\ref{sub:Governing_equations_central_object}, and those modelling self-gravity in~\S\ref{sub:Governing_equations_self_gravity}.

\subsection{Equations of motion for the disc}
\label{sub:Governing_equations_disc}

\renewcommand{\ur}{u_{r}} 
\newcommand{\up}{u_{\varphi}} 

\newcommand{\phis}{\phi_{\star}} 
\newcommand{\phid}{\phi_{d}} 

The standard way to describe the disc is to use the cylindrical coordinates $r$ and $\varphi$, the surface density $\Sigma$, the radial velocity $\ur$, the azimuthal velocity $\up$, the central object's gravitational potential $\phis$ and the disc's self potential $\phid$. But we prefer another set of variables introduced in \cite{TaggerPellat99} that makes the equations lighter. It simply replaces $r$ by ${ s = \ln{(r / R_{0})} }$, $\ur$ by ${ U = r \, \ur }$, $\up$ by ${ V = r \, \up }$ and $\Sigma$ by ${ \sigma = \ln{(\Sigma / \Sigma_{0})} }$ where \smash{$R_{0}$} and \smash{${\Sigma_{0}}$} are reference radii and surface densities. The Eulerian mass and momentum conservation equations then read
\begin{subequations}
    \small
    \label{eq:EoM_disc}
    \begin{align}
        & \!\!\!\! \partial_{t} \sigma + \frac{1}{r^{2}} \big[ \partial_{s} U + U \partial_{s} \sigma + \partial_{\varphi} V +  V \partial_{\varphi} \sigma \big] = 0 , \label{eq:EoM_disc_mass} \\
        & \!\!\!\! \partial_{t} U + \frac{U}{r^{2}} \big[ \partial_{s} U - U \big] + \frac{V}{r^{2}} \big[ \partial_{\varphi} U - V \big] = - \partial_{s} \big[ \phis + \phid \big] \! , \!\!\!\! \label{eq:EoM_disc_momentum_r} \\
        & \!\!\!\! \partial_{t} V + \frac{U}{r^{2}} \big[ \partial_{s} V - V] + \frac{V}{r^{2}} \big[ \partial_{\varphi} V + U \big] = - \partial_{\varphi} \big[ \phis + \phid \big] \! . \!\!\!\! \label{eq:EoM_disc_momentum_p}
    \end{align}
\end{subequations}

\subsection{Equations of motion for the central object}
\label{sub:Governing_equations_central_object}

\subsubsection{Barycentric reference frame}
\label{ssub:Governing_equations_central_object_barycentric}

\newcommand{\brs}{\br_{\star}} 

The barycentre of the \{disc + central object\} system is static, so the barycentric reference frame is inertial. Since there is no need for fictitious forces, the central object is only subject to the disc's gravity, and its equation of motion writes
\begin{equation}
    \label{eq:central_object_acceleration_barycentric}
    \rd_{tt}^{2} \, \brs = \iint\limits_{\text{disc}} \! \rd \br \,\,  \frac{G \, \Sigma(\br)}{||\br - \brs||^{2}} \, \frac{\br - \brs}{||\br - \brs||} ,
\end{equation}
where $\brs$ denotes the position of the central object and ${\rd \br}$ is a scalar surface element.

The gravitational potential of the central object is then 
\begin{equation}
    \label{eq:central_object_potential_barycentric}
    \phis (\br) = - \frac{G \Ms}{||\br - \brs||} ,
\end{equation}
where $G$ is the gravitational constant.

\subsubsection{Stellocentric reference frame}
\label{ssub:Governing_equations_central_object_stellocentric}

On the other hand, the benefit of the central object's reference frame is that the gravitational potential is constant,
\begin{equation}
    \label{eq:central_object_potential_stellocentric}
    \phis (\br) = - \frac{G \Ms}{r} .
\end{equation}

\newcommand{\phii}{\phi_{i}} 

\newcommand{\bas}{\ba_{\star}} 

However, since this reference frame is non-inertial, we need to add a fictitious force often called the `indirect term' to the disc's governing equations (although the prefix `the' should be avoided -- see \citealt{Crida+25a}). This force is uniform, so it derives from the `indirect potential'
\begin{equation}
    \label{eq:indirect_potential}
    \phii = \bas \bcdot \br ,
\end{equation}
where $\bas$ is the acceleration that the central object would have if we were in the barycentric reference frame.

To calculate this acceleration, we simply integrate the gravitational forces exerted by all the disc parcels,
\begin{equation}
    \label{eq:central_object_acceleration_stellocentric}
    \bas = \iint\limits_{\text{disc}} \! \rd \br \,\, \frac{G \, \Sigma(\br)}{r^{2}} \, \er ,
\end{equation}
where $\er$ is the radial unit vector.

\subsection{Self-gravity}
\label{sub:Governing_equations_self_gravity}

In both reference frames, the self-gravitational potential of the disc is simply
\begin{equation}
    \label{eq:self_gravitational_potential}
    \phid (\br) = - \iint\limits_{\text{disc}} \! \rd \br' \,\, \frac{G \, \Sigma (\br')}{||\br' - \br||} .
\end{equation}

\section{Equilibrium state}
\label{sec:Equilibrium}

\begin{figure}
    \centering
    \includegraphics[width = \linewidth]{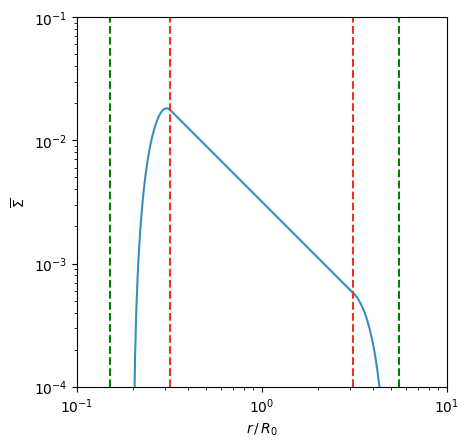}
    \caption{
        Typical equilibrium surface density profile of a disc with an inner cavity. Most disc simulations only model the power-law part of the disc (between the red vertical lines), but we shall see that this assumption activates the bracket modes. The only way to avoid those artefacts is to model the entire disc all the way from the inner green line in the central cavity to the outer green line well beyond the disc's outer edge, but this is prohibitively expensive.
    }
    \label{fig:Equilibrium}
\end{figure}

Our goal is to study the response of the system to tiny perturbations. But before we can do that, we need to define the equilibrium upon which infinitesimal perturbations might grow or propagate. This equilibrium is the same in both reference frames, and we shall denote equilibrium variables with an overline ${ \overline{\phantom{a}} }$.

\newcommand{\oSigma}{\overline{\Sigma}} 
\newcommand{\our}{\overline{u}_{r}} 
\newcommand{\oup}{\overline{u}_{\varphi}} 

\newcommand{\os}{\overline{\sigma}} 
\newcommand{\oU}{\overline{U}} 
\newcommand{\oV}{\overline{V}} 

\newcommand{\ophid}{\overline{\phi}_{d}} 

\newcommand{\obrs}{\overline{\br}_{\star}} 

\newcommand{\ri}{r_{\text{in}}} 
\newcommand{\ro}{r_{\text{out}}} 

In the absence of any perturbation, the central object is stationary (\smash{${\obrs = \mathbf{0} }$}) and the disc is circular (\smash{${\our = \bar{U} = \mathbf{0}}$}). 

The azimuthal velocity results from a balance between the central object's gravity, the centrifugal force and the disc's self-gravity, so we have \smash{${\oup = r \, \Omega(r) }$} with 
\begin{align}
    \Omega (r) &= \sqrt{\frac{G \Ms}{r^{3}} + \frac{1}{r} \rd_{r} \ophid} . \label{eq:rotation_frequency}
\end{align}

Finally, we need to describe the surface density profile. Following \cite{Crida+25b}, we assume that \smash{$\oSigma$} follows a power law, \smash{${ \oSigma (r) \propto r^{-p} }$}, of index ${p = 1.5}$ and ranging from ${ \ri = 0.32 R_{0} }$ to ${ \ro = 3.125 R_{0} }$. The proportionality coefficient, \smash{$\Sigma_{0}$}, is set by the mass of the disc relative to the central object, ${ \Md / \Ms }$, whose fiducial value is~$0.06$.

Note that using a power-law model implies that the modelling domain does not span the entire disc. Fig.~\ref{fig:Equilibrium} shows this visually: the boundaries of the domain are inside the disc. We shall see that this common assumption is at the origin of the bracket modes.

\section{Linear perturbation equations}
\label{sec:Linear_perturbation_equations}

Let us now perturb the background flow, \smash{${ \overline{f} }$}, with a perturbation, \smash{${ f' }$}, which we assume to be small, \smash{${ f' \ll \overline{f} }$}. The total flow is then \smash{${ f = \overline{f} + f' }$}, where ${ f }$ denotes any variable (density, velocity, potential, \textit{etc.}). We shall follow the same structure as in \S\ref{sec:Governing_equations} to show the linearised form of the equations.

\subsection{Equations of motion for the disc}
\label{sub:linear_perturbation_equations_disc}

If we neglect the quadratic terms of the form \smash{${ f' \times f' }$}, the disc's governing equations~\eqref{eq:EoM_disc} become
\begin{subequations}
    \label{eq:Linear_perturbation_equations}
    \begin{align}
        & \!\!\!\! \big( \partial_{t} + \Omega \, \partial_{\varphi} \big) \, \sigma' + \frac{1}{r^{2}} \left[ \partial_{s} U' + U' \, \rd_{s} \os + \partial_{\varphi} V' \right] = 0 , \label{eq:Linear_perturbation_equations_mass} \\
        & \!\!\!\! \big( \partial_{t} + \Omega \, \partial_{\varphi} \big) \, U' - 2 \Omega \, V' = - \partial_{s} \big[ \phis' + \phii' + \phid' \big] \! , \!\!\!\! \label{eq:Linear_perturbation_equations_momentum_r} \\
        & \!\!\!\! \big( \partial_{t} + \Omega \, \partial_{\varphi} \big) \, V' + \frac{\kappa^{2}}{2 \Omega} \, U' = - \partial_{\varphi} \big[ \phis' + \phii' + \phid' \big] . \label{eq:Linear_perturbation_equations_momentum_p}
    \end{align}
\end{subequations}
where \smash{${ \kappa = \sqrt{ \rd_{r} (r^{2} \, \Omega)^{2} / r^{3} } }$} is the local epicyclic frequency. Naturally, \smash{$\phii'$} is null in the barycentric reference frame and \smash{$\phis'$} is null in the stellocentric reference frame.

\newcommand{\hSigma}{\hat{\Sigma}} 
\newcommand{\hur}{\hat{u}_{r}} 
\newcommand{\hup}{\hat{u}_{\varphi}} 

\newcommand{\hs}{\hat{\sigma}} 
\newcommand{\hU}{\hat{U}} 
\newcommand{\hV}{\hat{V}} 

\newcommand{\hphis}{\hat{\phi}_{\star}} 
\newcommand{\hphii}{\hat{\phi}_{i}} 
\newcommand{\hphid}{\hat{\phi}_{d}} 

\newcommand{\C}{\mathcal{C}} 

Those equations are linear and none of the coefficients depend on time or azimuth, so we can look for Fourier-mode solutions of the form \smash{${ f(t, r, \varphi) = \widehat{f} (r) \, \e^{i (\varpi t - m \varphi)} }$}. Equations~\eqref{eq:Linear_perturbation_equations} simplify to
\begin{subequations}
    \label{eq:Fourier-transformed_equations}
    \begin{align}
        & \!\!\!\! i \C \, \hs + \frac{1}{r^{2}} \left[ \rd_{s} \hU + \hU \, \rd_{s} \os - i m \, \hV \right] = 0 , \label{eq:Fourier-transformed_equations_mass} \\
        & \!\!\!\! i \C \, \hU - 2 \Omega \, \hV = - \rd_{s} \big[ \hphis + \hphii + \hphid \big] \! , \!\!\!\! \label{eq:Fourier-transformed_equations_momentum_r} \\
        & \!\!\!\! i \C \, \hV + \frac{\kappa^{2}}{2 \Omega} \, \hU = i m \, \big[ \hphis + \hphii + \hphid \big] . \label{eq:Fourier-transformed_equations_momentum_p}
    \end{align}
\end{subequations}
where ${ \C = \varpi - m \Omega }$ is the Doppler-shifted frequency. We use the letter $\C$ to indicate that it is zero at corotation.


\subsection{Equations of motion for the central object}
\label{sub:linear_perturbation_equations_central_object}

\subsubsection{Barycentric reference frame}
\label{ssub:linear_perturbation_equations_central_object_barycentric}

\newcommand{\rsi}{r_{\star}'} 
\newcommand{\brsi}{\brs'} 
\newcommand{\xsi}{x_{\star}'} 
\newcommand{\ysi}{y_{\star}'} 

\newcommand{\hbrs}{\hat{\br}_{\star}} 

To model the linear dynamics of the central object, we follow \cite{Heemskerk+92}. First, we assume that \smash{${ \rsi \ll r }$} for every point \smash{$\br$} within the disc. This allows to expand the integrand in Eq.~\eqref{eq:central_object_acceleration_barycentric} in powers of \smash{${\rsi / r}$}:
\begin{small}
\begin{flalign}
    \!\!\!\!\!\! \rd_{tt}^{2} \, \brsi \approx \! \iint\limits_{\text{disc}} \! \rd \br \,\,  \frac{G \, \Sigma}{r^{2}} \bigg\{\er + 3 \cos{(\varphi)} \sin{(\varphi)} \left[\frac{\xsi}{r} \ey + \frac{\ysi}{r} \ex \right] \nonumber \\
    \pushright{ + \left[ 3 \cos{(\varphi)}^{2} - 1\right] \dfrac{\xsi}{r} \ex + \left[ 3 \sin{(\varphi)}^{2} - 1\right] \dfrac{\ysi}{r} \ey \bigg\} ,} \nonumber
\end{flalign}
\end{small}
\!\!\!\!where \smash{${\ex}$} is the unit vector of azimuth \smash{${ \varphi = 0 }$} and \smash{${\ey}$} that of azimuth \smash{${ \varphi = \pi / 2 }$}. $x$ and $y$ are the corresponding coordinates, and $\xsi$ and $\ysi$ the central object's displacement.

We then expand the surface density in powers of \smash{${ \Sigma' / \oSigma }$} and neglect the quadratic terms in \smash{${ (\rsi / r) \times (\Sigma' / \oSigma) }$}. This leaves five terms:
\begin{small}
\begin{flalign}
    \!\!\!\!\! \rd_{tt}^{2} \, \brsi \! \approx G \! \iint\limits_{\text{disc}} \! \rd \br \, \bigg( \frac{\oSigma \! + \! \Sigma'}{r^{2}} \er \! + \! \frac{\oSigma}{r^{2}}\bigg\{\frac{3}{2} \sin{(2 \varphi)} \! \left[ \! \frac{\xsi}{r} \ey \! + \! \frac{\ysi}{r} \ex \! \right] \nonumber \\
    \pushright{+ \left[ 3 \cos{(\varphi)}^{2} - 1\right] \dfrac{\xsi}{r} \ex + \left[ 3 \sin{(\varphi)}^{2} - 1\right] \dfrac{\ysi}{r} \ey \bigg\} \bigg) .} \nonumber
\end{flalign}
\end{small}

We then replace \smash{$\er$} by \smash{${ \cos{(\varphi)} \, \ex + \sin{(\varphi)} \, \ey }$} and \smash{$\Sigma'$} by \smash{${ \hSigma (r) \, \e^{i (\varpi t - m \varphi)} }$}. This allows us to compute the azimuthal integrals and get
\begin{equation}
    \nonumber
    \rd_{tt}^{2} \, \brsi \approx \delta_{m}^{\pm 1} \, \pi G \, \int_{\ri}^{\ro} \!\!\!\!\!\! \rd r \, \frac{\hSigma}{r} \left[ \ex \mp i \ey \right] \e^{i \varpi t} + \pi G \, \int_{\ri}^{\ro} \!\!\!\!\!\!\! \rd r \, \frac{\oSigma}{r^{2}} \brsi ,
\end{equation}
where ${ \delta_{m}^{\pm 1} }$ is a Kronecker symbol indicating that only modes ${m = \pm 1}$ can affect the central object's trajectory.

We recognise the equation of an inverted harmonic oscillator of natural frequency \smash{${ \mu = \scriptstyle \sqrt{ \pi G \int \! \rd r \, \oSigma / r^{2}} }$} and forced at frequency $\varpi$. The central object will respond at the same frequency, so we can write \smash{${ \brsi (t) = \hbrs \, \e^{i \varpi t} }$} to get
\begin{equation}
    \label{eq:Linear_perturbation_equations_central_object_location_barycentric}
    \hbrs = - \frac{\delta_{m}^{\pm 1} \, \pi G}{\varpi^{2} + \mu^{2}} \int_{\ri}^{\ro} \! \rd r \, \frac{\hSigma (r)}{r} \, \left[ \ex \mp i \ey \right]
\end{equation}

Finally, we expand the central object's potential~\eqref{eq:central_object_potential_barycentric} in powers of ${ \hbrs / \br }$ and truncate at first order to get
\begin{equation}
    \label{eq:Linear_perturbation_equations_central_object_potential_barycentric}
    \hphis (r) = \frac{\Omega_{\text{K}}^{2} (r)}{\varpi^{2} + \mu^{2}} \times \delta_{m}^{\pm 1} \, \pi G \, r \, \int_{\ri}^{\ro} \!\!\!\! \rd r' \, \frac{\hSigma (r')}{r'} ,
\end{equation}
where \smash{${ \Omega_{K} = \sqrt{G \Ms / r^{3}} }$} is the Keplerian frequency.

\subsubsection{Stellocentric reference frame}
\label{ssub:linear_perturbation_equations_central_object_stellocentric}

In the stellocentric reference frame, the central object is stationary, so its potential~\eqref{eq:central_object_potential_stellocentric} is unperturbed: \smash{${ \hphis = 0 }$}. The indirect potential~\eqref{eq:indirect_potential}, on the other hand, can be perturbed by ${ m  = \pm 1 }$ modes. Following similar steps as for the barycentric frame, we get
\begin{equation}
    \label{eq:Linear_perturbation_equations_indirect_potential}
    \hphii (r) = \delta_{m}^{\pm 1} \, \pi G \, r \, \int_{\ri}^{\ro} \!\!\!\! \rd r' \, \frac{\hSigma (r')}{r'} .
\end{equation}
If necessary, the full derivation can be found in our following paper.

\subsection{Self-gravity}
\label{sub:linear_perturbation_equations_self_gravity}

Since the surface density perturbation is of the form \smash{${ \hSigma (r) \, \e^{i (\varpi t - m \varphi)} }$}, we can simplify Eq.~\eqref{eq:self_gravitational_potential} to
\begin{equation}
    \label{eq:Linear_perturbation_equations_self_gravity_potential}
    \hphid (r) = - G \int_{\ri}^{\ro} \rd r' \, K_{m} \left( \frac{r'}{r} \right) \, \hSigma (r') ,
\end{equation}
where
\begin{equation}
    \label{eq:Gravitational_kernel}
    K_{m} (x) = x \int_{0}^{2 \pi} \rd \varphi \,\, \frac{\cos{(m \varphi)}}{\sqrt{1 + x^{2} - 2 x \cos{(\varphi)}}}
\end{equation}
is the gravitational kernel. It has a singularity in~${ x = 1 }$, so we regularise it by adding an~$\epsilon^{2}$ term inside the square root, with ${ \epsilon = 10^{-2} }$.

\section{Bracket modes in the barycentric reference frame}
\label{sec:Bracket_modes_barycentric}

Let us now focus on a set of modes discovered by \cite{Sanchez+25} and which occur when the boundaries of the simulation domain are inside the disc, \textit{i.e.} when \smash{${\oSigma (\ri) }$} and/or \smash{${\oSigma (\ro) }$} are not null. We call them `bracket' modes because they arise from the bracket side of an integration by parts. For the sake of clarity, we first study those modes without self-gravity in \S\ref{sub:Bracket_modes_barycentric_mathematical_description}. Then we show in \S\ref{sub:Bracket_modes_barycentric_sg} that unless the disc is extremely massive, self-gravity does not affect them. We only consider the ${ m = +1 }$ case because the ${m = -1}$ case is equivalent.

\subsection{Mathematical description}
\label{sub:Bracket_modes_barycentric_mathematical_description}

\renewcommand{\L}{\mathcal{L}} 

In the absence of self-gravity and indirect terms, we can reduce the linear momentum equations~(\ref{eq:Fourier-transformed_equations_momentum_r}, \ref{eq:Fourier-transformed_equations_momentum_p}) to
\begin{subequations}
    \label{eq:Bracket_modes_barycentric_momentum_equations}
    \begin{align}
        & i \C \, \hU - 2 \Omega \, \hV = - \rd_{s} \hphis , \!\!\!\!\! \label{eq:Bracket_modes_barycentric_momentum_equations_U} \\
        & i \C \, \hV \! + \! \frac{\kappa^{2}}{2 \Omega} \, \hU = i \, \hphis , \label{eq:Bracket_modes_barycentric_momentum_equations_V}
    \end{align}
\end{subequations}
quickly leading to \addtocounter{equation}{-1}
\begin{subequations}
    \nonumber
    \begin{align}
        & \hU = i \, \frac{2 \Omega \, \hphis - \C \, \rd_{s} \hphis}{\L^{2}} , \\
        & \hV = \frac{\frac{\kappa^{2}}{2 \Omega} \rd_{s} \hphis - \C \, \hphis}{\L^{2}} ,
    \end{align}
\end{subequations}
where ${ \L^{2} = \kappa^{2} - \C^{2} }$. We use the letter $\L$ to indicate that it is zero at the Lindblad resonances. 

\newcommand{\Lo}{\L_{\text{o}}} 

Note that since there is no pressure nor self-gravity, \smash{${ \kappa = \Omega }$}. Therefore, there is no inner \smash{${ m = 1 }$} Lindblad resonance and we can replace \smash{${ \L^{2} }$} by \smash{${ - \varpi \Lo }$} where \smash{${ \Lo = \varpi - 2 \Omega }$} represents the outer Lindblad resonance.

\newcommand{\hrsx}{\hat{r}_{\star, x}} 

We can also introduce \smash{${ \hrsx = \hbrs \bcdot \ex }$} in order to write \smash{${ \hphis = - r \Omega^2 \hrsx }$} and \smash{${ \rd_{s} \hphis = + 2 r \Omega^{2} \hrsx }$}. With those notations, the momentum equations yield
\begin{subequations}
    \label{eq:Bracket_mode_barycentric_velocities}
    \begin{align}
        \hU &= + \, i \,\, r \Omega \, \frac{2 \Omega}{\Lo} \, \hrsx , \label{eq:Bracket_mode_barycentric_velocities_U} \\
        \hV &= -\frac{1}{2} \, r \Omega \, \frac{2 \Omega}{\Lo} \, \hrsx . \label{eq:Bracket_mode_barycentric_velocities_V}
    \end{align}
\end{subequations}

We then inject those formulas into the mass conservation equation~\eqref{eq:Fourier-transformed_equations_mass} to get the density perturbation,
\begin{equation}
    \label{eq:Bracket_mode_baryentric_surface_density}
    \hSigma = - \frac{\Omega}{\C} \, \rd_{r} \left( \frac{2 \Omega}{\Lo} \, \oSigma \right) \hrsx .
\end{equation}

Now remember that the equation of motion for the star~\eqref{eq:Linear_perturbation_equations_central_object_location_barycentric} can be written
\begin{equation}
    \nonumber
    - \varpi^{2} \, \hrsx = \pi G \, \int_{\ri}^{\ro} \!\!\!\!\!\! \rd r \, \frac{\hSigma}{r} + \mu^{2} \, \hrsx .
\end{equation}
We can integrate by parts the \smash{${\hSigma}$} term and develop the \smash{${\mu^{2}}$} term to get the dispersion relation,
\begin{small}
\begin{flalign}
    \!\!\!\!\!\!\!\!\!\!\!\!\!\!\!\!\!\!\!\!\!\!\!\!\!\!\!\!\!\!\!\!\!\!\!\!\!\!\!\!\!\! \varpi^{2} = \pi G \, \bigg\{ \! \left[ \frac{\Omega}{\C} \frac{2 \Omega}{\Lo} \frac{\oSigma}{r} \right]_{\ri}^{\ro} - \int_{\ri}^{\ro} \!\!\!\!\!\! \rd r \, \frac{2 \Omega}{\Lo} \frac{\oSigma}{r} \, \rd_{r} \frac{\Omega}{\C} \nonumber \\ 
    \pushright{\displaystyle + \int_{\ri}^{\ro} \!\!\!\!\!\! \rd r \, \left[ \frac{\Omega}{\C} \frac{2 \Omega}{\Lo} - 1 \right] \frac{\oSigma}{r^{2}} \bigg\} .}
\end{flalign}
\end{small}

\newcommand{\tr}{\tilde{r}} 
\newcommand{\tvarpi}{\tilde{\varpi}} 
\newcommand{\tOmega}{\tilde{\Omega}} 
\newcommand{\tC}{\tilde{\C}} 
\newcommand{\tLo}{\tilde{\L}_{\text{o}}} 
\newcommand{\tSigma}{\tilde{\Sigma}} 

\newcommand{\tri}{\tilde{r}_{\text{in}}} 
\newcommand{\tro}{\tilde{r}_{\text{out}}} 

We adimensionalise this equation using \smash{${ \ro }$} as a reference lengthscale, \smash{${ 1 / \Omega (\ro) }$} as a reference timescale, and \smash{${ \Ms / \pi \ro^{2} }$} as a reference surface density. We get
\begin{small}
\begin{flalign}
    \!\!\!\!\!\!\!\!\!\!\!\!\!\!\!\!\!\!\!\!\!\!\!\!\!\!\!\!\!\!\!\!\!\!\!\!\!\!\!\!\!\!\! \tvarpi^{2} = \bigg\{ \! \left[ \frac{\tOmega}{\tC} \frac{2 \tOmega}{\tLo} \frac{\tSigma}{\tr} \right]_{\tri}^{\tro} - \int_{\tri}^{\tro} \!\!\!\!\!\! \rd \tr \, \frac{2 \tOmega}{\tLo} \frac{\tSigma}{\tr} \, \rd_{\tr} \frac{\tOmega}{\tC} \nonumber \\ 
    \pushright{\displaystyle + \int_{\tri}^{\tro} \!\!\!\!\!\! \rd \tr \, \left[ \frac{\tOmega}{\tC} \frac{2 \tOmega}{\tLo} - 1 \right] \frac{\tSigma}{\tr^{2}} \bigg\} .}
\end{flalign}
\end{small}

\begin{figure}
    \centering
    \includegraphics[width = \linewidth]{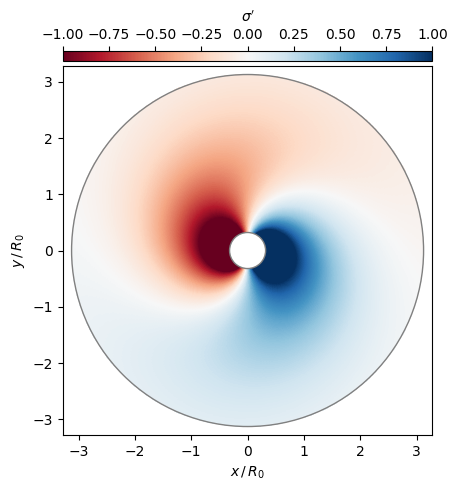}
    \caption{
        2D representation of the relative density perturbation \smash{${ \sigma' (r, \varphi, t) = \hs(r) \, \e^{i (\varpi t - \varphi)} }$} induced by the unstable barycentric bracket mode in our fiducial disc. It takes the form of a loosely wound one-armed spiral. Remember that the maximum amplitude of the eigenvector is a gauge choice, it does not have any physical meaning. What matters are the relative amplitudes between the perturbations to different variables and at different radii. We therefore normalise the mode such that \smash{${ \hrsx / R_{0} = 3.0 }$}. The eigenfrequency and the density perturbation were computed by our new linear analysis code LAADS -- more on that in \S\ref{sub:Bracket_modes_barycentric_sg}. $t$ is set to zero.
    }
    \label{fig:2D}
\end{figure}

Finally, we assume that \smash{${ \tvarpi \! \ll \! 1 }$}. This means that the pattern frequency of the mode is much smaller than the Keplerian frequency \textit{everywhere in the disc}. This explains why we use \smash{$\ro$} instead of \smash{${R_{0}}$} to adimensionalise the previous equation. Since \smash{${ \tOmega / \tC \! \approx \! 2 \tOmega / \tLo \! \approx \! - 1 }$}, we can simplify the expression to
\begin{equation}
    \label{eq:Barycentric_eigenfrequency}
    \tvarpi^{2} = \left[ \frac{\tSigma}{\tr} \right]_{\tri}^{\tro} + \mathcal{O} \left( \frac{\tvarpi \, \tSigma}{\tr^{2}} \right) .
\end{equation}
This shows that if the boundaries of the model are inside the disc, as shown in Fig.~\ref{fig:Equilibrium}, the system exhibits two `bracket' modes. For the power-law disc described in~\S\ref{sec:Equilibrium}, the bracket is negative so one of those modes is unstable\footnote{We explain the physical mechanism of this instability in \S\ref{sec:Barycentric_bracket_modes_mechanism}.} and the other one damped. We show in Fig.~\ref{fig:2D} the relative density perturbation induced by the unstable one. If the bracket was positive, however, the two modes would be marginally stable. One would be prograde and the other one retrograde.

Eq.~\eqref{eq:Barycentric_eigenfrequency} also shows that those modes are artificial. Indeed, their growth rate depends on $\ri$ and~$\ro$, which characterise the model but not the physical system.

\subsection{Addition of self-gravity}
\label{sub:Bracket_modes_barycentric_sg}

\cite{Crida+25a} showed that for a model to be consistent, it needs to account for self-gravity whenever it accounts for the reflex motion of the central object. Therefore, we must investigate the effect of self-gravity on the bracket modes.

\begin{figure}
    \centering
    \includegraphics[width = \linewidth]{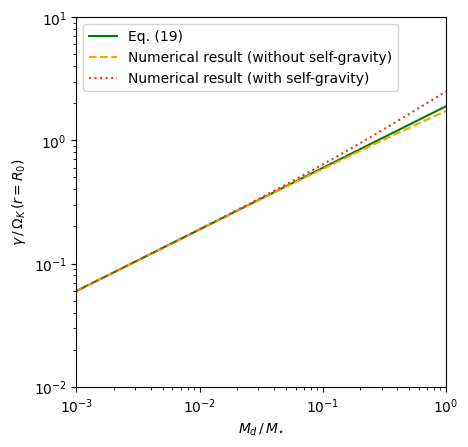}
    \vspace{-1.5 \baselineskip}
    \caption{
        Growth rate \smash{${ \gamma = - \text{Im} (\varpi) }$} of the unstable bracket mode of the barycentric equations. We use the disc model of \S\ref{sec:Equilibrium} with and without self-gravity. The solid green line represents the analytical prediction from Eq.~\eqref{eq:Barycentric_eigenfrequency}, which we expect to be accurate in low-mass discs. The dashed orange line and the dotted red line represent the numerical predictions from LAADS without and with self-gravity. The match between the green and orange lines validates the analytical prediction, while the match between the orange and red lines indicates that self-gravity has little impact.
    }
    \label{fig:Effect_of_self_gravity_barycentric}
\end{figure}

To do so, we use a new linear analysis code called LAADS and which we will present in a future publication. Since this code has not gone through the peer-review process yet, our reader is entitled to doubt its correctness. In order to build trust, we invite our reader to compare the green and orange curves of Fig.~\ref{fig:Effect_of_self_gravity_barycentric}. The green curve represents the growth rate of the unstable bracket mode as predicted by Eq.~\eqref{eq:Barycentric_eigenfrequency}, while the orange curve represents LAADS' prediction. The excellent match between those two predictions suggests that our analytic work and our numerical work are both correct.

The red curve represents the numerical prediction when self-gravity is turned on. Even for a heavy disc with \smash{${ \Md / \Ms = 0.1 }$}, self-gravity only increases the growth rate by 7\%. We therefore conclude that self-gravity only has a minor effect on the barycentric bracket modes.

\section{Bracket modes in the stellocentric reference frame}
\label{sec:Bracket_modes_stellocentric}

Let us now reproduce the same analysis in the stellocentric reference frame. We shall see that the stellocentric bracket modes are qualitatively different from the barycentric ones.

\subsection{Mathematical description}
\label{sub:Bracket_modes_stellocentric_mathematical_description}

In the stellocentric frame, $\hphis$ is null but $\hphii$ isn't. Since it is proportional to $r^{1}$ instead of $r^{-2}$, the velocity \mbox{perturbations} become
\begin{subequations}
    \label{eq:Bracket_mode_stellocentric_velocities}
    \begin{align}
        \hU &= i \, r \, \frac{\varpi - 3 \Omega}{\Lo} \, \frac{\hphii}{r \, \varpi} , \label{eq:Bracket_mode_stellocentric_velocities_U} \\
        \hV &= r \, \frac{\varpi - 3 \Omega / 2}{\Lo} \, \frac{\hphii}{r \, \varpi} , \label{eq:Bracket_mode_stellocentric_velocities_V}
    \end{align}
\end{subequations}
and the surface density perturbation
\begin{equation}
    \label{eq:Bracket_mode_stellocentric_surface_density}
    \!\!\! \hSigma = - \frac{1}{2} \frac{\Omega}{\C} \left[ \frac{\varpi}{\Omega} \frac{2 \Omega}{\Lo} \frac{\oSigma}{r \, \Omega} + \frac{\varpi - 3 \Omega}{\Omega} \rd_{s} \left( \frac{2 \Omega}{\Lo} \frac{\oSigma}{r \, \Omega} \right) \right] \frac{\hphii}{r \, \varpi} . \!\!\!
\end{equation}

Substituting this into Eq.~\eqref{eq:Linear_perturbation_equations_indirect_potential} gives the dispersion relation,
\begin{small}
\begin{equation}
    \nonumber
    \varpi = - \frac{1}{2} \, \pi G \, \int_{\ri}^{\ro} \!\!\!\! \rd r \, \frac{\Omega}{\C} \left[ \frac{\varpi}{\Omega} \frac{2 \Omega}{\Lo} \frac{\oSigma}{r^{2} \, \Omega} + \frac{\varpi \! - \! 3 \Omega}{\Omega} \rd_{r} \left( \frac{2 \Omega}{\Lo} \frac{\oSigma}{r \, \Omega} \right) \right] \! ,
\end{equation}
\end{small}
\!\!\!\!which adimensionalises to
\begin{small}
\begin{equation}
    \nonumber
    \tvarpi = - \frac{1}{2} \, \int_{\tri}^{\tro} \!\!\!\! \rd \tr \, \frac{\tOmega}{\tC} \left[ \frac{\tvarpi}{\tOmega} \frac{2 \tOmega}{\tLo} \frac{\tSigma}{\tr^{2} \, \tOmega} + \frac{\tvarpi \! - \! 3 \tOmega}{\tOmega} \rd_{\tr} \left( \frac{2 \tOmega}{\tLo} \frac{\tSigma}{\tr \, \tOmega} \right) \right] \! .
\end{equation}
\end{small}

Finally, just like in \S\ref{sec:Bracket_modes_barycentric}, let us assume that \smash{${ \tvarpi \ll 1 }$}. In this limit, the dispersion relation reduces~to
\begin{equation}
    \label{eq:bracket_modes_stellocentric}
    \tvarpi = \frac{3}{2} \, \left[ \frac{\tSigma}{\tr \, \tOmega} \right]_{\tri}^{\tro} + \mathcal{O} \left( \frac{\tvarpi \, \tSigma}{\tr^{2} \, \tOmega^{2}} \right) .
\end{equation}
Just like in the barycentric reference frame, if the disc extends beyond the boundaries of the simulation, the system exhibits artificial modes. However, this time there is only one mode and it is marginally stable (\textit{i.e.} its growth rate is exactly zero). In the power-law disc of~\S\ref{sec:Equilibrium}, it is retrograde. Indeed, its pattern frequency is negative.

Note that even though the stellocentric mode is stable, it still pollutes simulations by creating a fake reservoir of energy. Indeed, non-linear interactions can transfer energy from some physical modes to the bracket mode, and release this energy back to different physical modes. The amplitudes of some physical modes are therefore underestimated while the amplitude of others are overestimated.

\subsection{Addition of self-gravity}
\label{sub:Bracket_modes_stellocentric_sg}

\begin{figure}
    \centering
    \includegraphics[width = \linewidth]{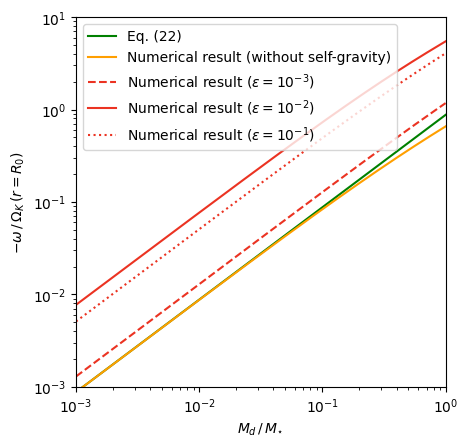}
    \caption{
        Same as Fig.~\ref{fig:Effect_of_self_gravity_barycentric}, but for the pattern frequency \smash{${ \omega \! = \! \Re (\varpi) }$} of the stellocentric bracket mode. The three red lines show that the multiplication factor induced by self-gravity depends on the softening parameter~$\epsilon$ in a non-trivial way.
    }
    \label{fig:Effect_of_self_gravity_stellocentric}
\end{figure}

Fig.~\ref{fig:Effect_of_self_gravity_stellocentric} shows that, contrary to the barycentric case, self-gravity has a strong effect on the stellocentric bracket mode. Indeed, it multiplies the pattern frequency by a finite factor whose value depends non-monotonically on the softening parameter $\epsilon$. The discrepancy induced by this factor between the numerical results with and without self-gravity in low-mass discs remains to be understood. We simply note that since the multiplicative factor is always positive and the mode retrograde, the self-gravitational boost cannot activate any corotation or Lindblad resonance.

\section{Discussion and conclusion}
\label{sec:Discussion_conclusion}

We investigated the modes discovered by \cite{Sanchez+25}. We provide the first analytical description of those modes' pattern frequency and growth rate in the main text, and the first physical explanation of their mechanism in appendix~\ref{sec:Barycentric_bracket_modes_mechanism}. Both clearly show that `bracket' modes are artificial (we propose this name because they arise from the bracket side of an integration by parts). Indeed, their growth rate and pattern frequency depend on the location of the edges of the simulation domain. The subtlety is that those artefacts are not numerical. Indeed, they are already present in the analytical equations, and can be studied analytically.

Nevertheless, they can pollute simulations just as much as numerical artefacts. In particular, they drive eccentricity at the global scale \citep{Sanchez+25}, so they could affect disc evolution and structure studies. Figure~\ref{fig:2D} shows that they also induce large-scale spirals, so they could mislead experts investigating the substructures observed by ALMA and the VLT \citep{Bae+23}. Thirdly, they redistribute mass within the disc (Eqs.~\ref{eq:Bracket_mode_baryentric_surface_density} and~\ref{eq:Bracket_mode_stellocentric_surface_density}), so they could bias planet migration experiments. Finally, they are stronger in massive discs (Eq.~\ref{eq:Barycentric_eigenfrequency}), so they could have major effects in simulations of gravitationally unstable discs, for instance on clump migration.

We discover that, contrary to what \cite{Sanchez+25} suggested, bracket modes are not necessarily unstable. This depends on the reference frame in which one models the system. This surprising and concerning finding is due to the fact that while the internal equations are frame-invariant (as long as we add the relevant fictitious forces), the boundaries of the domain are frame-dependent. Indeed, they are static in the barycentric reference frame but tied to the central object in the stellocentric reference frame.\footnote{In fact, the stability of the bracket modes also depends on which equations we use to model the reflex motion of the central object. We can either model it `dynamically' using Newton's second law, or `conservatively' by explicitly imposing centre-of-mass stationarity. We chose to only present the `dynamic' equations in this paper, to avoid accumulating algebra and diluting our message. But our explorations in~\S\ref{sec:Conservative_equations} suggest that the `conservative' equations do not exhibit any bracket mode, regardless of stability.}

This hints at the only robust way to avoid those artefacts. Since they appear when the surface density of the unperturbed disc is non-zero at the edge of the simulation domain, one should model the entire disc all the way from the central cavity to the outer edge. But such simulations are prohibitively expensive, and young protoplanetary discs may not have an inner cavity anyway \citep{FiorellinoSomigliana26}.

Looking at the mathematical structure of our equations, it seems likely that similar issues will affect all systems where the disc is subject to a large-scale, slowly-evolving perturbation. This includes fly-bys, outer binary companions, photo-evaporation, ... We therefore advise authors to do their due diligence when making claims regarding eccentricity, spirals or planet migration in those systems. One way to do that is to repeat the analysis in the other reference frame. In parallel, we also advise readers to exercise scepticism when evaluating such claims.

Indeed, analytical artefacts are harder to recognise and avoid than numerical artefacts. And in the present case, dependence on the reference frame makes things even harder. This leads us to suggest that our community may not have the right tools to model the reflex motion of the central object yet. We hope that our work will motivate the development of such a tool.

In the meantime, it would be good to study how the bracket instability develops in the non-linear regime. Indeed, it does not require pressure and therefore cannot saturate through fluid pathways such as shocks or turbulence.

It would also be interesting to explore the links between our analysis and the variational methods used by \cite{Toomre69}, \cite{Kalnajs71} and \cite{Narayan+87}. In particular, how do the brackets of Eqs.~\eqref{eq:Barycentric_eigenfrequency} and~\eqref{eq:bracket_modes_stellocentric} relate to the action and angular momentum flux shown by those authors to control disc stability? In fact, once we account for the reflex motion and the angular momentum of the central object, does this flux still control stability?

In our next paper, we show that the four degrees of freedom brought by the central object (\smash{$x_{\star}$}, \smash{$y_{\star}$}, \smash{$\dot{x}_{\star}$} and \smash{$\dot{y}_{\star}$}) are sufficient to activate at least seven instabilities, some of which have already been seen in the literature.

\section*{Acknowledgments}

NM acknowledges financial support from the Centre National d’\'Etudes Spatiales (CNES - France) within the framework of the Hera mission, from the COST action CA22133 `PLANETS', and from the French government through the \smash{${\text{UCA}^{\text{JEDI}}}$} `Investments in the Future' project managed by the National Research Agency (ANR) with the reference number \smash{ANR-15-IDEX-01}. All authors acknowledge financial support from the `Programme National Astronomie - Astrophysique' (PNAA) and `Programme National de Plan\'etologie' (PNP) of CNRS/INSU co-funded by CEA and CNES.

\section*{Data Availability}

The data and numerical codes underlying this article were produced by the authors. They will be shared on reasonable request to the corresponding author.

\section*{References}
\bibliographystyle{mnras}
\bibliography{main}

@ARTICLE{Rees88,
       author = {{Rees}, Martin J.},
        title = "{Tidal disruption of stars by black holes of {}10$^{6}$-{}10$^{8}$ solar masses in nearby galaxies}",
      journal = {\nat},
         year = 1988,
        month = jun,
       volume = {333},
       number = {6173},
        pages = {523-528},
          doi = {10.1038/333523a0},
       adsurl = {https://ui.adsabs.harvard.edu/abs/1988Natur.333..523R}
}

@ARTICLE{Fragile+26,
       author = {{Fragile}, P. Chris and {Ingram}, Adam and {Musoke}, Gibwa and {Ogilvie}, Gordon I.},
        title = "{Tilted, Warped, and Eccentric Disks}",
      journal = {\ssr},
         year = 2026,
        month = mar,
       volume = {222},
       number = {2},
          eid = {25},
        pages = {25},
          doi = {10.1007/s11214-026-01280-z},
archivePrefix = {arXiv},
       eprint = {2603.02993},
 primaryClass = {astro-ph.HE},
       adsurl = {https://ui.adsabs.harvard.edu/abs/2026SSRv..222...25F}
}

@INPROCEEDINGS{Bae+23,
       author = {{Bae}, J. and {Isella}, A. and {Zhu}, Z. and {Martin}, R. and {Okuzumi}, S. and {Suriano}, S.},
        title = "{Structured Distributions of Gas and Solids in Protoplanetary Disks}",
    booktitle = {Protostars and Planets VII},
         year = 2023,
       editor = {{Inutsuka}, S. and {Aikawa}, Y. and {Muto}, T. and {Tomida}, K. and {Tamura}, M.},
       series = {Astronomical Society of the Pacific Conference Series},
       volume = {534},
        month = jul,
        pages = {423},
          doi = {10.48550/arXiv.2210.13314},
archivePrefix = {arXiv},
       eprint = {2210.13314},
 primaryClass = {astro-ph.EP},
       adsurl = {https://ui.adsabs.harvard.edu/abs/2023ASPC..534..423B}
}

@ARTICLE{Adams+89,
       author = {{Adams}, Fred C. and {Ruden}, Steven P. and {Shu}, Frank H.},
        title = "{Eccentric Gravitational Instabilities in Nearly Keplerian Disks}",
      journal = {\apj},
         year = 1989,
        month = dec,
       volume = {347},
        pages = {959},
          doi = {10.1086/168187},
       adsurl = {https://ui.adsabs.harvard.edu/abs/1989ApJ...347..959A}
}

@ARTICLE{Heemskerk+92,
       author = {{Heemskerk}, M.~H.~M. and {Papaloizou}, J.~C. and {Savonije}, G.~J.},
        title = "{Non-linear development of M = 1 instabilities in a self-gravitating gaseous disc}",
      journal = {\aap},
         year = 1992,
        month = jul,
       volume = {260},
       number = {1-2},
        pages = {161-174},
       adsurl = {https://ui.adsabs.harvard.edu/abs/1992A&A...260..161H}
}

@ARTICLE{Crida+25b,
       author = {{Crida}, Aur{\'e}lien and {Baruteau}, Cl{\'e}ment and {Gonzalez}, Jean-Fran{\c{c}}ois and {Masset}, Fr{\'e}d{\'e}ric and {Segretain}, Paul and {Griveaud}, Philippine and {M{\'e}heut}, H{\'e}lo{\"\i}se and {Lega}, Elena},
        title = "{The reflex instability: exponential growth of a large-scale m = 1 mode in astrophysical discs}",
      journal = {The Open Journal of Astrophysics},
         year = 2025,
        month = aug,
       volume = {8},
          eid = {115},
        pages = {115},
          doi = {10.33232/001c.143230},
archivePrefix = {arXiv},
       eprint = {2508.07859},
 primaryClass = {astro-ph.EP},
       adsurl = {https://ui.adsabs.harvard.edu/abs/2025OJAp....8E.115C}
}

@ARTICLE{Sanchez+25,
       author = {{Sanchez}, M. and {Paardekooper}, S. and {van der Marel}, N. and {Ben{\'\i}tez-Llambay}, P. and {Mulders}, G.~D.},
        title = "{New migration patterns in high planet─star mass ratio systems in discs with low viscosity}",
      journal = {\aap},
         year = 2025,
        month = nov,
       volume = {703},
          eid = {A169},
        pages = {A169},
          doi = {10.1051/0004-6361/202555041},
archivePrefix = {arXiv},
       eprint = {2509.15719},
 primaryClass = {astro-ph.EP},
       adsurl = {https://ui.adsabs.harvard.edu/abs/2025A&A...703A.169S}
}

@ARTICLE{TaggerPellat99,
       author = {{Tagger}, M. and {Pellat}, R.},
        title = "{An accretion-ejection instability in magnetized disks}",
      journal = {\aap},
         year = 1999,
        month = sep,
       volume = {349},
        pages = {1003-1016},
          doi = {10.48550/arXiv.astro-ph/9907267},
archivePrefix = {arXiv},
       eprint = {astro-ph/9907267},
 primaryClass = {astro-ph},
       adsurl = {https://ui.adsabs.harvard.edu/abs/1999A&A...349.1003T}
}

@ARTICLE{Crida+25a,
       author = {{Crida}, Aur{\'e}lien and {Baruteau}, Cl{\'e}ment and {Griveaud}, Philippine and {Lega}, Elena and {Masset}, Fr{\'e}d{\'e}ric and {B{\'e}thune}, William and {Fang}, David and {Gonzalez}, Jean-Fran{\c{c}}ois and {M{\'e}heut}, H{\'e}lo{\"\i}se and {Morbidelli}, Alessandro and {Gerosa}, Fabiola and {Kloster}, Dylan and {Marques}, L{\'e}a and {Miniussi}, Alain and {Minker}, Kate and {Pichierri}, Gabriele and {Segretain}, Paul},
        title = "{On inertial forces (indirect terms) in problems with a central body}",
      journal = {The Open Journal of Astrophysics},
         year = 2025,
        month = jul,
       volume = {8},
          eid = {84},
        pages = {84},
          doi = {10.33232/001c.141682},
archivePrefix = {arXiv},
       eprint = {2506.23331},
 primaryClass = {astro-ph.EP},
       adsurl = {https://ui.adsabs.harvard.edu/abs/2025OJAp....8E..84C}
}

@ARTICLE{FiorellinoSomigliana26, 
AUTHOR={Fiorellino, Eleonora  and Somigliana, Alice },        
TITLE={The accretion process on protostars},      
JOURNAL={Frontiers in Astronomy and Space Sciences},       
VOLUME={Volume 13},
YEAR={2026}, 
URL={https://www.frontiersin.org/journals/astronomy-and-space-sciences/articles/10.3389/fspas.2026.1819945},
DOI={10.3389/fspas.2026.1819945},
ISSN={2296-987X}
}

@ARTICLE{Crida+07,
       author = {{Crida}, A. and {Morbidelli}, A. and {Masset}, F.},
        title = "{Simulating planet migration in globally evolving disks}",
      journal = {\aap},
         year = 2007,
        month = jan,
       volume = {461},
       number = {3},
        pages = {1173-1183},
          doi = {10.1051/0004-6361:20065870},
archivePrefix = {arXiv},
       eprint = {astro-ph/0608660},
 primaryClass = {astro-ph},
       adsurl = {https://ui.adsabs.harvard.edu/abs/2007A&A...461.1173C}
}

@ARTICLE{Toomre69,
       author = {{Toomre}, Alar},
        title = "{Group Velocity of Spiral Waves in Galactic Disks}",
      journal = {\apj},
         year = 1969,
        month = dec,
       volume = {158},
        pages = {899},
          doi = {10.1086/150250},
       adsurl = {https://ui.adsabs.harvard.edu/abs/1969ApJ...158..899T}
}

@ARTICLE{Kalnajs71,
       author = {{Kalnajs}, Agris J.},
        title = "{Dynamics of Flat Galaxies. I}",
      journal = {\apj},
         year = 1971,
        month = jun,
       volume = {166},
        pages = {275},
          doi = {10.1086/150957},
       adsurl = {https://ui.adsabs.harvard.edu/abs/1971ApJ...166..275K}
}

@ARTICLE{Narayan+87,
       author = {{Narayan}, Ramesh and {Goldreich}, Peter and {Goodman}, Jeremy},
        title = "{Physics of modes in a differentially rotating system - Analysis of the shearing sheet}",
      journal = {\mnras},
         year = 1987,
        month = sep,
       volume = {228},
        pages = {1-41},
          doi = {10.1093/mnras/228.1.1},
       adsurl = {https://ui.adsabs.harvard.edu/abs/1987MNRAS.228....1N}
}

\vspace{0.5 \baselineskip}
\appendix

\section{A. Physical picture of the instability mechanism}
\label{sec:Barycentric_bracket_modes_mechanism}

The goal of this appendix is to leverage our analytical understanding of the bracket modes into physical understanding. To keep things simple, we focus on the unstable barycentric bracket mode. We provide a `comic-book' version of our explanation in Fig.~\ref{fig:Physical_picture_of_the_unstable_barycentric_bracket_mode} and advise our reader to read each portion in parallel with the corresponding panels.

\subsection{A.1. Setup and start of the feedback loop}
\label{sub:Mechanism_setup}

Let us assume that initially, the central object is not exactly at the centre of the disc. Since it is closer/further to some parts of the disc than usual, this displacement induces a potential perturbation. This is shown in panel~1 and explains the \textit{r.h.s.} of Eqs.~\eqref{eq:Bracket_modes_barycentric_momentum_equations}.

\begin{figure*}
    \centering
    \includegraphics[width = 0.88 \textwidth]{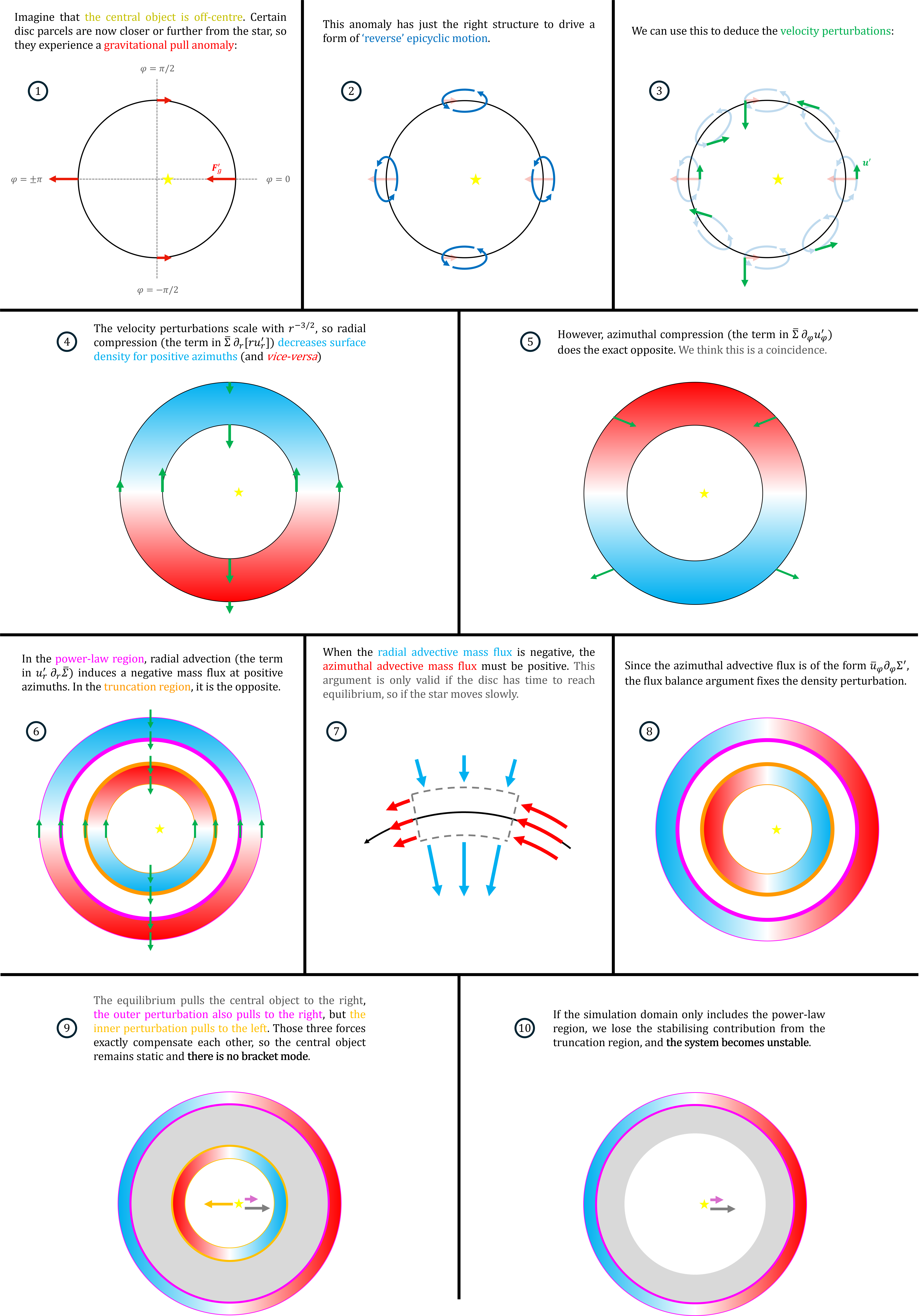}
    \caption{
        `Comic book' explanation of the instability mechanism of the barycentric bracket mode. The panels should be read from left to right and from top to bottom.
    }
    \label{fig:Physical_picture_of_the_unstable_barycentric_bracket_mode}
\end{figure*}

The disc must respond to this forcing. Note that since we use a Fourier transform and not a Laplace transform, we can only study the secular response and not the transient response. Without this crucial piece of context, one might misinterpret Fig.~\ref{fig:Physical_picture_of_the_unstable_barycentric_bracket_mode}.

\subsection{A.2. Velocity perturbations}
\label{sub:Mechanism_velocity_perturbation}

Since the perturbation is a force, the disc material will accelerate and respond through velocity perturbations.

The factor \smash{${i/2}$} between \smash{$\hU$} and \smash{$\hV$} in Eqs.~\eqref{eq:Bracket_mode_barycentric_velocities} suggests that the disc parcels describe epicycles. However, panel~2 shows that the parcels follow the ellipses in the wrong direction. Indeed, the total azimuthal velocity is maximal at apogee and minimal at perigee. Panel~3 displays the velocity perturbations associated to those `fake~epicycles'.

\subsection{A.3. Density perturbations}
\label{sub:Mechanism_density_perturbation}

Fluid motion redistributes mass within the disc, so the velocity perturbations induce density perturbations. In principle, this can occur through radial advection, azimuthal advection, radial compression, or azimuthal compression. However, Eq.~\eqref{eq:Fourier-transformed_equations_mass} shows that considering our symmetries, only four mass fluxes may be active at the linear order: azimuthal advection of the density perturbation by the background velocity, radial advection of the background density by the velocity perturbation, radial compression of the background density by the velocity perturbation, and azimuthal compression of the background density by the velocity perturbation.

\subsubsection{A.3.1. Compression does not affect the mass distribution}
\label{ssub:Mechanism_density_perturbation_no_compression}

By injecting Eqs.~\eqref{eq:Bracket_mode_barycentric_velocities} into Eq.~\eqref{eq:Fourier-transformed_equations_mass}, we see that in the small frequency limit, \smash{${\tvarpi \ll 1}$}, the radial and azimuthal compression terms exactly compensate each other. Panels~4 and~5 show this visually.

Note that this exact compensation is not indicative of a hidden symmetry of reverse epicyclic motion. It is simply a mathematical oddity of the \smash{${ r^{-2} }$} scaling of \smash{${ \hphis }$}. Indeed, compression does play a role in the stellocentric reference frame where \smash{${ \hphii }$} scales with \smash{${ r^{1} }$}.

\subsubsection{A.3.2. Radial advection}
\label{ssub:Mechanism_density_perturbation_radial_advection}

The radial advection flux is proportional to \smash{${ \rd_{r} \oSigma }$}, which is usually negative. But if we consider a disc with an inner cavity, the radial flux can be inverted in the `truncation' region that connects the cavity to the rest of the disc. This is shown in panel~6.

\subsubsection{A.3.3. Flux balance and azimuthal advection}
\label{ssub:Mechanism_density_perturbation_azimuthal_advection}

At low frequency, Eq.~\eqref{eq:Fourier-transformed_equations_mass} indicates that the radial and azimuthal advection terms must also compensate each other. This is shown in panel~7. 

Panel~8 then shows that, since the azimuthal advective flux is proportional to \smash{${ \rd_{\varphi} \hSigma }$}, we can deduce the density perturbation caused by the velocity perturbations.

Crucially, since the radial flux takes opposite signs in the inner and outer discs, mass accumulates on opposite sides of the disc in those two regions.

\subsection{A.4. Closing the feedback loop}
\label{sub:Mechanism_closing_the_feedback_loop}

Now that we know the mass distribution in the disc, we can compute the gravitational pull exerted by the mass asymmetry on the central object. Since we know that the system is unstable, we expect to find that the central object is being pulled away from the origin.

\subsection{A.4.1. Break-down of the gravitational pull}
\label{sub:Mechanism_closing_the_feedback_loop_three_contributions}

There are three contributions to the gravitational pull of the disc on the central object, all represented in panel~9:
\begin{itemize}
    \vspace{-0.5 \baselineskip}
    \item[(i)] The central object is slightly off-centre, so it is closer to one side of the disc than the other. Since gravity scales with \smash{${ r^{-2} }$}, this means that the equilibrium mass distribution attracts the central object to the right. This explains the \smash{${ \mu^{2} }$} in Eq.~\eqref{eq:Linear_perturbation_equations_central_object_location_barycentric}. 
    \vspace{-0.5 \baselineskip}
    \item[(ii)] The density perturbation in the outer disc also pulls the central object to the right.
    \vspace{-0.5 \baselineskip}
    \item[(iii)] But the density perturbation in the inner disc pulls to the left.  
\end{itemize}

\subsection{A.4.2. No instability in a full disc}
\label{sub:Mechanism_closing_the_feedback_loop_no_instability_in_full_disc}

Equation~\eqref{eq:Barycentric_eigenfrequency} shows that if we model the entire disc all the way from the inner cavity to the outer edge, then those three contributions exactly compensate each other (whatever the surface density profile). And since the net force experienced by the star is null, the system is stable.

\subsection{A.4.3. Instability in a power-law disc}
\label{sub:Mechanism_closing_the_feedback_loop_instability_in_power-law_disc}

Panel~10 shows that if -- like many grid-based simulations -- we only model the power-law part of the disc, then we lose the stabilising contribution from the truncation region. Since the star experiences a net force to the right, the system is unstable. But this instability is clearly artificial.

\section{B. No bracket modes with the conservative equations}
\label{sec:Conservative_equations}

We claimed in a footnote at the end of the main part of the paper that when one models the position of the central object through centre-of-mass conservation rather than gravitational attraction, there is no bracket mode. The goal of the present appendix is to justify that claim. 

To do so, we shall run the same analysis as in~\S\ref{sec:Bracket_modes_barycentric} and~\S\ref{sec:Bracket_modes_stellocentric} but for the conservative equations, and show that the new dispersion relations do not have any low-frequency root. We start with the barycentric reference frame in~\S\ref{sub:Conservative_barycentric} and end with the stellocentric frame in~\S\ref{sub:Conservative_stellocentric}.

\subsection{B.1. Barycentric reference frame}
\label{sub:Conservative_barycentric}

\subsubsection{B.1.1. Governing equations}
\label{ssub:Conservative_barycentric_governing_equations}

In \S\ref{ssub:Governing_equations_central_object_barycentric}, we found the position of the central object by using Newton's laws of motion and universal gravitation. Since this requires solving a differential equation in time, we call this the `dynamic' approach. The `conservative' approach is simpler in that it only involves an algebraic equation. The idea is to leverage the fact that the centre of mass of the \{disc + central object\} is at the origin of the reference frame. By integrating the mass moments in the disc, we get
\begin{equation}
    \label{eq:central_object_position_barycentric_conservative}
    \brs = - \frac{1}{\Ms} \iint\limits_{\text{disc}} \! \rd \br \,\,  \Sigma(\br) \, \br .
\end{equation}

This is the approach used by the Eulerian code FARGO-2D1D \citep{Crida+07}.

\subsubsection{B.1.2. Linearised equations}
\label{ssub:Conservative_barycentric_linearised_equations}

Applying the same linearisation and Fourier-transform procedures as in \S\ref{ssub:linear_perturbation_equations_central_object_barycentric} to Eq.~\eqref{eq:central_object_position_barycentric_conservative} rather than Eq.~\eqref{eq:central_object_acceleration_barycentric} gives
\begin{equation}
        \label{eq:linear_perturbation_equations_barycentric_conservative_position}
        \hbrs = - \frac{\delta_{m}^{\pm 1} \, \pi}{\Ms} \int_{\ri}^{\ro} \!\!\!\! \rd r \, r^{2} \, \hSigma \, \left[ \ex \mp i \, \ey \right].
\end{equation}
The corresponding potential perturbation is
\begin{equation}
    \label{eq:linear_perturbation_equations_barycentric_conservative_potential}
    \hphis (r) = \frac{\Omega^{2} (r)}{\Omega_{0}^{2}} \times \delta_{m}^{\pm 1} \, \pi G \, r \, \int_{\ri}^{\ro} \!\!\!\! \rd r' \, \frac{\Omega_{0}^{2}}{\Omega^{2} (r')} \, \frac{\hSigma (r')}{r'} ,
\end{equation}
where $\Omega_{0}$ is a reference frequency. Its value has no importance, its only role is to make the connections and differences between Eqs.~\eqref{eq:Linear_perturbation_equations_central_object_potential_barycentric} and~\eqref{eq:linear_perturbation_equations_barycentric_conservative_potential} more apparent.

\subsubsection{B.1.3. Dispersion relation}
\label{ssub:Conservative_barycentric_dispersion_relation}

Since we still have \smash{${ \hphis = - r \Omega^2 \hrsx }$}, Equation~\eqref{eq:Bracket_mode_baryentric_surface_density} still stands. Injecting it in Eq.~\eqref{eq:linear_perturbation_equations_barycentric_conservative_position} gives
\begin{equation}
    \nonumber
    \int_{\ri}^{\ro} \!\!\!\!\!\! \rd r \, r^{2} \, \frac{\Omega}{\C} \, \rd_{r} \left[ \frac{2 \Omega}{\Lo} \oSigma \right] = \frac{\Ms}{\pi} ,
\end{equation}
which adimensionalises to
\begin{equation}
    \nonumber
    \int_{\tri}^{\tro} \!\!\!\!\!\! \rd \tr \, \tr^{2} \, \frac{\tOmega}{\tC} \, \rd_{\tr} \left[ \frac{2 \tOmega}{\tLo} \tSigma \right]  = 1 .
\end{equation}

\subsubsection{B.1.4. Interpretation}
\label{ssub:Conservative_barycentric_interpretation}

In the limit of small frequencies, ${ \tvarpi \ll 1 }$, the dispersion relation simplifies to
\begin{equation}
    \nonumber
    \int_{\tri}^{\tro} \!\!\!\!\!\! \rd \tr \, \tr^{2} \, \rd_{\tr} \tSigma = 1 .
\end{equation}
Unless $\ri$ and $\ro$ are chosen very carefully, this equality is never verified. This indicates that the small-frequency limit is inconsistent.

We postulate that the conservative barycentric equations do not allow for bracket modes. Indeed, all the modes we found by solving the linear problem numerically with LAADS were artificial (either their corotation radius or their outer Lindblad radius fell within 0.1\% of a grid point).

\subsection{B.2. Stellocentric reference frame}
\label{sub:Conservative_stellocentric}

\subsubsection{B.2.1. Governing equations}
\label{ssub:Conservative_stellocentric_governing_equations}

\cite{Adams+89} compute the position of the barycentre in the stellocentric reference frame,
\begin{equation}
    \nonumber
    \br_{b} = \frac{1}{\Ms + \Md} \, \iint\limits_{\text{disc}} \! \rd \br \,\, r \, \Sigma (\br) \, \er ,
\end{equation}
and point out that the acceleration of the star in the barycentric reference frame is the opposite of the acceleration of the barycentre in the stellocentric frame,
\begin{equation}
    \label{eq:central_object_acceleration_stellocentric_conservative}
    \bas = - \frac{\rd^{2} \, \br_{b}}{\rd t^{2}} .
\end{equation}
This is all we need to compute the indirect potential~\eqref{eq:indirect_potential}.

\subsubsection{B.2.2. Linearised equations}
\label{ssub:Conservative_stellocentric_linearised_equations}

The usual linearisation process leads to
\begin{small}
    \begin{equation}
        \label{eq:linear_perturbation_equations_stellocentric_conservative_indirect_potential}
        \hphii (r) = \frac{\Ms}{\Ms + \Md} \times \delta_{m}^{\pm 1} \, \pi G \, r \, \int_{\ri}^{\ro} \!\!\!\! \rd r' \, \frac{\varpi^{2}}{\Omega^{2} (r')} \, \frac{\hSigma (r')}{r'} .
    \end{equation}
\end{small}

\subsubsection{B.2.3. Dispersion relation}
\label{ssub:Conservative_stellocentric_dispersion_relation}

Since \smash{$\hphii$} is still proportional to \smash{$r^{1}$}, Equation~\eqref{eq:Bracket_mode_stellocentric_surface_density} remains valid. Injecting it into Eq.~\eqref{eq:linear_perturbation_equations_stellocentric_conservative_indirect_potential} leads to
\begin{flalign}
    \!\!\!\!\!\!\!\!\!\!\!\!\!\!\!\!\!\!\!\!\!\!\!\!\!\!\!\!\!\!\!\!\!\!\!\!\!\!\! \frac{1}{\varpi} = - \frac{1}{2} \, \frac{\pi}{\Ms + \Md} \, \int_{\ri}^{\ro} \!\!\!\! \rd r' \, \frac{\Omega}{\C} \bigg[ \frac{\varpi}{\Omega} \frac{2 \Omega}{\Lo} \frac{\oSigma}{\Omega / r'} \nonumber \\
    \pushright{\displaystyle + \frac{\varpi - 3 \Omega}{\Omega / r'^{2}} \rd_{s'} \left( \frac{2 \Omega}{\Lo} \frac{\oSigma}{r'' \, \Omega} \right) \bigg] ,} \nonumber
\end{flalign}
which adimensionalises to
\begin{flalign}
    \!\!\!\!\!\!\!\!\!\!\!\!\!\!\!\!\!\!\!\!\!\!\!\!\!\!\!\!\!\!\!\!\!\!\!\!\!\!\! \frac{2}{\tvarpi} = - \frac{\Ms}{\Ms + \Md} \, \int_{\tri}^{\tro} \!\!\!\! \rd \tr' \, \frac{\tOmega}{\tC} \bigg[ \frac{\tvarpi}{\tOmega} \frac{2 \tOmega}{\tLo} \frac{\tSigma}{\tOmega / \tr'} \nonumber \\
    \pushright{\displaystyle + \frac{\tvarpi - 3 \tOmega}{\tOmega / \tr'^{2}} \rd_{s'} \left( \frac{2 \tOmega}{\tLo} \frac{\tSigma}{\tr'' \, \tOmega} \right) \bigg] ,} \nonumber
\end{flalign}

\subsubsection{B.2.4. Interpretation}
\label{ssub:Conservative_stellocentric_interpretation}

This time, in the limit of small frequencies, the dispersion relation reduces to
\begin{equation}
    \nonumber
    \frac{1}{\tvarpi} = \frac{3}{2} \, \frac{\Ms}{\Ms + \Md} \, \int_{\tri}^{\tro} \!\!\!\! \rd \tr' \, \tr'^{3} \, \rd_{\tr'} \left( \frac{\tSigma}{\tr \, \tOmega} \right) .
\end{equation}
The \textit{r.h.s.} is not large, so $\tvarpi$ is not small, and the small-frequency limit is inconsistent. We tried to optimise \smash{${ \ri / \ro }$} and \smash{${ \Md / \Ms }$} to maximise the right-hand side, but could not recover any bracket mode numerically. Therefore, we postulate that just like the conservative barycentric equations, the conservative stellocentric equations do not exhibit any bracket mode.

In both reference frames, the reason there is a difference between the `dynamic' and `conservative' approaches is that centre-of-mass conservation only derives from the Navier-Stokes equations in closed systems. Here the boundaries of the domain are inside the disc and the radial velocity perturbations are not null, so mass can leave the domain, and centre-of-mass conservation does not follow from the governing equations~\eqref{eq:EoM_disc}. 

Note that neither approach is `correct'. Indeed, they both fail to conserve mass. The only safe way to make the two approaches equivalent and the two reference frames equivalent is to model the entire disc all the way from the inner cavity to the outer edge.

\vspace{+1 \baselineskip}

\end{document}